# Attomole enantiomeric discrimination of small molecules using an achiral SERS reporter and chiral plasmonics.

*Shailendra K Chaubey[a,\*], Rahul Kumar[a], Paula Laborda[a], Martin Kartau[a], Victor Tabouillot[a], Oleksiy Lyutakov[b], Nikolaj Gadegaard[c], Affar S. Karimullah[a], Malcolm Kadodwala[a,\*]*

[a]School of Chemistry, Joseph Black Building, University of Glasgow, Glasgow, G12 8QQ, UK

[b] *Department of Solid-State Engineering, University of Chemistry and Technology, 16628 Prague, Czech Republic*

[c]School of Engineering, Rankine Building, University of Glasgow, Glasgow, G12 8QQ, UK

Keywords: Chirality, SERS, Plasmonics

## Abstract

Biologically important molecules span a size range from very large biomacromolecules, such as proteins (> 11 kDa) to small metabolite molecules (~0.1 kDa). Consequently, spectroscopic techniques which can detect and characterize the structure of inherently chiral biomolecules over this range of scale at the ≤ femtomole level are necessary to develop novel biosensing and diagnostic technologies. Nanophotonic platforms uniquely enable chirally sensitive structural characterisation of biomacromolecules at this ultrasensitive level. However, they are less successful at achieving the same level of sensitivity for small chiral molecules, with > nanomole typical. This poorer performance can be attributed to the optical response of the platform being sensitive to a much larger volume of the near field than is occupied by the small molecule. Here we show that by combining chiral plasmonic metasurfaces with Raman reporters, which can detect changes in electromagnetic environment at molecular dimensions, chiral discrimination can be achieved for attomole ($10^{-18}$M) quantities of a small molecule, the amino acid cysteine. The signal-to-noise, and hence ultimate sensitivity, of the measurement can be further improved by combining the metasurfaces with gold achiral nanoparticles. This "indirect" enantiomeric detection is ≥ 9 orders of magnitude more sensitive than strategies relying on monitoring the Raman response of target chiral molecules directly. Given the generic nature of the phenomenon, this study provides a framework for developing novel technologies for detecting a broad spectrum of small biomolecules, which would be useful tools in the field of metabolomics.

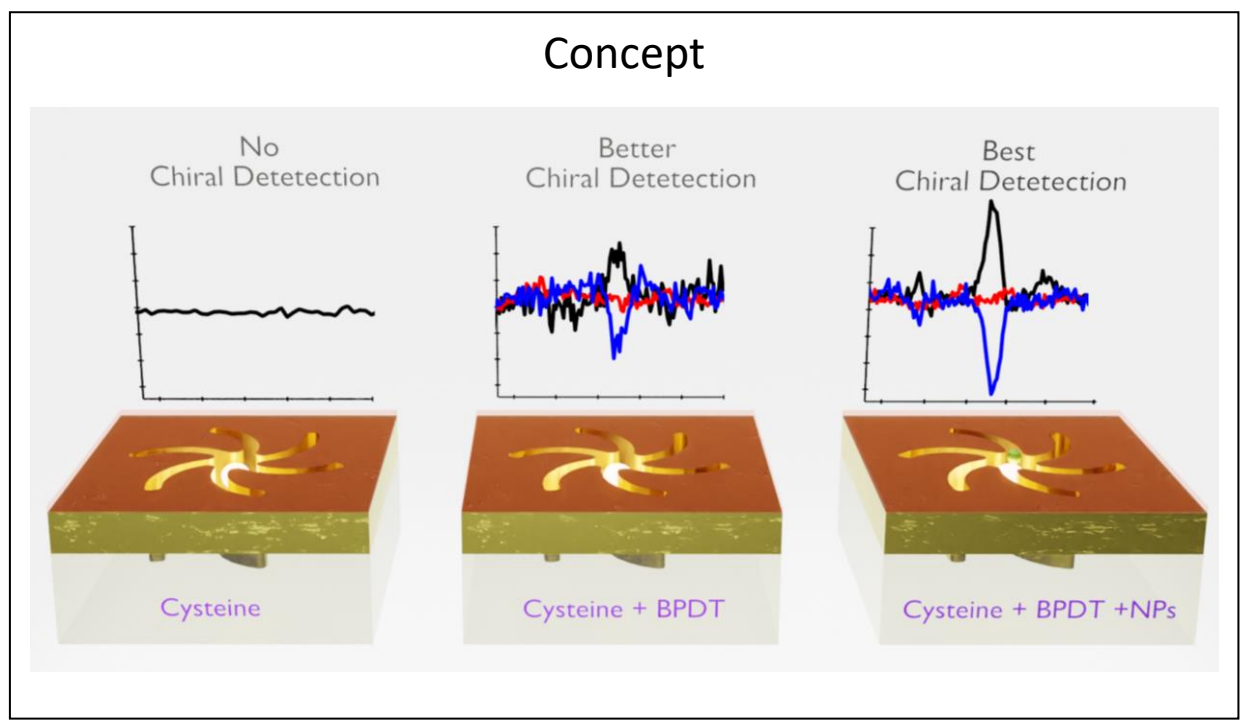

## Introduction

Chirally sensitive spectroscopic methods, based on monitoring the dichroic interactions of circularly polarised light, are important biophysical tools providing rapid characterisation of biomolecules. However, the inherently small magnitudes of dichroic phenomena severely limit sensitivity, typically to the micromole (μM) levels, which prevents exploitation of these powerful tools in applications such as point of care diagnostics and biosensing which require < femtomole (fM) detection levels. Nanophotonic platforms provide novel paradigms for achieving this level of ultrasensitive chiral detection[1-5]. They can allow chirality of molecules to be detected without relying on monitoring the inherently weak dichroic absorption and scattering used by conventional chiroptical spectroscopy. Instead, the detection strategy relies on detecting asymmetric changes in the local electromagnetic (EM) environments of enantiomorphic pairs of structures induced by the introduction of chiral molecules.[6-9] Since, asymmetries are observed even when the near field is non-resonant with the electronic excitations of biomolecules, which appear in the near UV. This allows gold (plasmonic) nanostructures which display resonances in the visible and near IR (NIR) regions to be used. The current strategies of how these platforms are utilized for chiral sensing have some intrinsic weaknesses. The changes in the EM environment are highly localized to regions of space occupied by the chiral (bio)molecules. Thus, this reduces the asymmetries observed in far field optical response based on light scattering which are sensitive to the larger global EM near field environment. Consequently, only relatively large peptides and proteins ( ≥10 kDa), which occupy relatively large volumes of the near field, can be studied using methods based on a combination of chiral nanostructures and extinction spectra-based measurements. This gap in capability narrows the range of applications. The detection of small chiral molecules would be an important tool for monitoring metabolites and pharmaceutical drugs.

Here we demonstrate a novel paradigm of nanophotonic based chiral sensing which enables the enantiomeric discrimination of < monolayer (equivalent to attomole) concentrations of a small chiral molecule, cysteine (0.121 kDa). The philosophy of the proposed chiral sensing concept is that the changes induced by a chiral small molecule in the EM environment are detected by a second non-chiral (achiral) reporter molecule. Specifically, we demonstrate that using a common achiral Raman reporter, biphenyl-4,4'-dithiol (BPDT), to monitor EM environments at a molecular scale, chiral sensitive detection at the attomole level can be achieved using enantiomorphic pairs of metasurfaces composed of shuriken shaped nanocavities. Due to the small size of cysteine and BPDT molecules <monolayer quantities are invisible to plasmonic far field reflectance measurements. In addition, the small scattering cross section of cysteine means that that < monolayer quantities are invisible to Raman, in contrast to the significant signals generated by BPDT. The co-deposition of gold nanoparticles (NPs) enhances the intensity of the BPDT signal and hence improves detection signal-to-noise. The reported phenomenon is a ≥ 9 orders of magnitude more sensitive than previously reported Raman based strategies that utilise metamaterials to achieve chiral discrimination[10, 11]. These previous studies have provided chiral discrimination of either drop cast films or liquids which have thicknesses in the micron range. These previous studies rely on strategies which directly monitor the sometimes-weak Raman response of the chiral molecule. By leveraging a molecular reporter with a strong Raman response sensitivity is enhanced. The proposed concept is ideally suited for translation into analytical technology. It exploits Raman scattering, a ubiquitous tool for chemical sciences, and chiral metasurfaces formed on polycarbonate templates manufactured using high throughput injection moulding.

**Result and Discussion**

Chiral gold metasurfaces used in this study consisted of periodic square arrays of six-armed chiral "shuriken" shaped nanocavities, figure 1 (a). The metasurfaces were formed by depositing 100 nm thick gold films on nanopatterned polycarbonate templates, which have been described in detail elsewhere.[7, 12, 13] Similar Au metasurfaces have been used in chiral sensing applications in which far field extinction[4, 8, 13-17] and luminescence[8, 9] spectra have been used to monitor asymmetries in the EM environments. Reflection and Optical Rotation Dispersion (ORD) spectra from enantiomorphic films collected in air display multiple resonances. As expected, similar reflection spectra are obtained from both enantiomorphs, which display a broader resonance in the 550-560 nm spectral range and another sharper resonance in 700-760nm spectra, figure 1 (d). Equal and opposite ORD spectra are obtained from the enantiomorphic pairs, both have resonances in the 550-600 nm range with bisignate line shapes (See supplementary information S1).

BPDT was chosen for this study for two reasons: 1) it has a very large Raman scattering cross-section[18-20], thus making it an effective reporter molecule; and 2) because the two SH functional groups enable it to chemisorb to the metasurface and also facilitates the binding of Au NP. Self-assembled monolayers of: L-/D-/Racemic-cysteine only, BPDT only or mixed -/D-/Racemic-cysteine and BPDT are deposited from aqueous solutions on to the chiral metasurfaces. Both the cysteines and BPDT form chemisorbed moieties through the cleavage of the S-H and formation of strong S-Au bonds[21-23]. The cysteine is expected to form a chemisorbed zwitter ion on the surface[21, 22, 24]. An important question to address is whether the adsorbed cysteine and BPDT moieties are inter-mixed, or phase separate into individual domains. Based on the relative polarities of cysteine (polar) and BPDT (non-polar), and hence the hydrophilic and hydrophobic properties of each respectively, we propose that this would drive phase separation into individual domains. Reflectance spectra collected after the deposition of the mixed cysteine / DPBT, figure 1 (d), and all other SAMs (See supplementary information S2) show no significant shifts relative to those collected prior to functionalisation.

The adsorption of cysteine on a range of Au surfaces has been extensively studied with a variety of techniques. The accepted coverages of cysteine within known surface structures are of the order of ~$2.0\times10^{14}$ molecules / cm$^2$ [23, 25]. Thus, within the ~100 μm$^2$ field of view sampled by the Raman microscopy there are $\leq 3\times10^{-18}$ moles ($\leq$ 3 attomoles) of cysteine.

Gold NPs of either 20 or 30 nm diameter were deposited on to the cysteine / BPDT SAM functionalised metasurfaces. SEM images, figure 1 (b) & (c), collected after deposition show that the surface density of 20 nm is greater than that of 30 nm diameter NPs. Also 20 diameter NPs have a greater heterogeneity in spatial distribution, than the larger diameter NPs. The greater spatial heterogeneity displayed by the 20 nm NPs is attributed to the smaller diameter allowing adsorption within the "arms" of the shuriken structure. In contrast the 30 nm particles are comparable in size to the arm cavity. Both 20 and 30 nm NPs have a propensity to bind to the side wall of the nanocavity. This behaviour would be consistent with a thermodynamic driver to maximise the co-ordination of the NP to free thiols groups of the BPDT, since it allows the particle to bind both to the surfaces of the walls and bottom of the cavity simultaneously. Consequently, this results in 30 nm NP favouring binding sites in the regions at the entrance of the arm cavity. As would be expected reflectance spectra collected after NP deposition show significant red shifts in the plasmonic resonances, figure 1 (e) & (f), with the largest observed for the 20 nm diameter NPs.

Both in the presence and absence of NP there was no observable asymmetry in the reflection or ORD spectra collected from enantiomorphic metasurfaces which had been functionalised with SAMs containing L-or D- cysteine. Thus, no enantiomeric detection for cysteine is possible using extinction spectra-based measurements.

Collection of SERS spectra were attempted from all the SAM functionalised surfaces, both metasurface and for references unstructured gold, in the absences and presence of gold NP, figure 1 (g). Each displayed spectrum is the average of spectra taken from 3 different arrays of the same enantiomorph, and each of these is the average of 5 different measurements. A SERS response was only observed from metasurfaces which had been functionalised with SAM contained BPDTs. The observed SERS spectra are solely dominated by bands associated with BPDT, which is consistent with the much weaker Raman scattering cross-section of cysteine. The spectra in this work are very similar to those obtained previously from BPDT immobilized on either Au nanoparticles or commercial SERS (Klarite) substrates.[26] The characteristic spectrum of BPDT has five dominant peaks, at 1030, 1084, 1200, 1285 and 1590 cm$^{-1}$, which we have labelled A, B, C, D, and E respectively. The vibrational assignments of these modes are given in Table 1. In the absence of NPs, the Raman response of mixed BPDT / cysteine layers are ~30% of that from pure BPDT layers. This implies that the mixed SAMs contain ~70% cysteine. Adding 30 nm and 20 nm particle enhances the Raman signal significantly. An additional peak at 1130 cm$^{-1}$ is visible in Raman spectra of NP containing layers. A similar feature has been observed in previous studies involving Au NPs and has been attributed to a Fermi resonance.[19]

The introduction of the NPs enhances the BPDT Raman signals from the mixed layers, with the presence of 20 and 30 nm NPs increasing the response by ~3.8 and 2.8 times respectively. Although the overall enhancement factors are similar, the much smaller concentration of the 30 nm NP makes them significantly more effective per particle.

To account for sample-to-sample variations in the absolute intensities of Raman signal, and hence aid comparison, all spectra have been normalised to the peak E, figure 2 (a)-(i). These spectra clearly demonstrate asymmetric Raman responses from enantiomeric pairs of metasurfaces when adsorbed L- or D- cysteine. With peak B showing significant intensity asymmetries between enantiomorphs, the sign of which depends on the handedness of the adsorbed cysteine. In the absence of L- or D-cysteine or for racemic cysteine there is no asymmetry in the Raman response from the enantiomorphic pairs of metasurfaces (See supplementary information S3). In figure 3 (a) – (d) are difference spectra derived from the subtracting, left minus right, metasurface data. The greater Raman response generated by the introduction of the NP, improves the signal to noise of the difference spectra. However, only the 30 nm NPs result in an increase in the level of asymmetry, while 20 NPs give a smaller asymmetry to that observed from just the mixed layers.

The central premise of this study is that the experimentally observed asymmetries in the Raman responses from enantiomorphic pairs are the result of asymmetric changes in the EM environments caused by the presence of the L- / D-cysteine. To validate this hypothesis, numerical EM simulations have been performed using the finite element method implemented on the COMSOL Multiphysics platform. Simulations were performed using an idealised model shuriken structure (see supplementary information S4, S5 and S6) and the chiral layer. The model consists of a "perfect" shuriken structure covered by a 5 nm dielectric layer (refractive index = 1.44) which can be assigned a Pasteur factor ($\xi$) to parameterise its chiroptical response. A cysteine monolayer would have a thickness of ~ 0.5 nm, however due to computational constraints a greater thickness had to be used. This should mean that a smaller $\xi$ should be used in the model than is physically realistic, to off-set the larger volume occupied by the chiral dielectric. Further simplifications were made for the models incorporating the NP. Only a single NP was included in these models, at a single site, which based on the SEM images, figure 1 (b) & (c) is a reasonable assumption for 30 nm particle, but unrealistic for the 20 nm case. A mid-point in the arm cavity was the location for the 20 nm NP. While for the 30 nm case the NP was placed at the entrance of the arm cavity.

Estimates of the relative intensities of the Raman signal can be derived from the EM simulations using the volume averaged magnitudes of electric fields within the chiral layer of the models. The magnitude of a Raman response of a particular band ($I_i$) is:

$$I_i \propto |E_{ex}|^2 |E_i|^2$$

where $|E_{ex}|$ and $|E_i|$ are the volume averaged field magnitudes at the excitation (632.8 nm) and Raman band wavelength. Consequently, the level of asymmetry (g) observed in the difference spectra was derived using

$$g = \left( \frac{|E_L^{Ex}|^2 |E_L^B|^2}{|E_L^{Ex}|^2 |E_L^E|^2} - \frac{|E_R^{Ex}|^2 |E_R^B|^2}{|E_R^{Ex}|^2 |E_R^E|^2} \right)$$

Where $\left|E_{L(R)}^{Ex}\right|$ are the magnitude of the electric fields at the excitation wavelength for left (right) metasurfaces. While $\left|E_{L(R)}^{B(E)}\right|$ are the magnitude of the electric at the wavelength of the B (E) peak of the left (right) enantiomorphs. The asymmetries were calculated with data from two simulations that used $\xi$ values of ($\pm$)$1.7 \times 10^{-4}$ and ($\pm$)$2.4 \times 10^{-3}$, table 2. The simulations replicate the relative sizes of the asymmetry, in the absence and presence of 20 and 30 nm NPs. However, the simulations can only mimic the absolute magnitude of the asymmetry with unrealistically large $\xi$ values ($2.4 \times 10^{-3}$). We speculate that this is due to the exclusion from the simulation model of the dipole associated with the zwitterionic state. Previous work has shown that the charge states of chiral molecules strongly influence how they alter the EM environments of chiral plasmonic structures. [13] This dependency may arise because the presence of a charge molecule with a dipole moment modifies the level of capacitive coupling within the structure. Another origin for the discrepancy between simulation and experiment is that the rough morphologies of real metasurfaces are not accounted for in the simulations. This means that simulations using idealised structural models do not account for hot spots of enhanced chiral asymmetry ("superchiral hotspots")[27].

Although the models that include NP are very crude, the electric field maps, figure 4, derived from them may provide some insight into why 30 nm particles are more effective at enhancing sensitivity. A potential factor that would affect the relative magnitudes of asymmetries is corelated to the size of the gap regions the NPs make with the side walls of the shuriken cavities. The larger 30 nm NPs can adopt geometries which on average have smaller gaps, thus more intense fields, than the 20 nm particles. Another parameter which may reduce the relative size of the asymmetries for the 20 nm diameter NPs is the heterogeneity in the position of the NP. This results in the particles occupying a broader range of EM environments which could smear out the asymmetries[8].

## Conclusions

This study demonstrates the potential advantages of an "indirect" detection strategy when utilising nanophotonic platforms for ultrasensitive chiral discrimination. Monolayers of cysteine are invisible to both Raman and extinction-based measurements. However, using a Raman probe, one can monitor the local EM environment at the molecular scale. This allows asymmetries in the EM environments of enantiomorphic pairs of nanocavities caused by the presence of an enantiomer of a small molecule to be measured. The combination of Raman spectroscopy with metasurfaces manufactured with high throughput techniques, make the phenomenon reported ideal for exploitation for novel technologies required for metabolomics and the pharmaceutical sectors.

## Methods

***Fabrication of Chiral Metasurface***: The polycarbonate templates were prepared by injection moulding. A silicon substrate was cleaned and coated with 80 PMMA (polymethyl methacrylate) to perform the electron beam lithography. The pattern was transferred using electron beam writing (Elvacite 2041, Lucite International) operating at 100 kV. After the lithography, the pattern was developed by immersion in a mixture of methyl isobutyl ketone and isopropyl alcohol. This pattern was transferred on 300 µm thick nickel sim by electroplating on the Si master slide. This nickel sim is mounted on custom made setup attached with a fully automated injection moulding machine (Engel Victory Tech 28 tons) to produce polymer slide using polycarbonate (Makrolon DP2015) as feedstock. The plastic sample were coated with gold using electron beam evaporation to make the chiral metafilm.

***Sample preparation***: The SAM formation on the metasurface was achieved using solution phase deposition. Briefly, shuriken metasurfaces were first sonicated in ethanol for 5 minutes, followed by plasma cleaning for 2 minutes at 100 W. 0.5 mg/ml solution of cysteine and biphenyl-4,4'-dithiol are prepared in water and methanol respectively and mixed together in a 1:1 ratio if required. The metasurfaces were immersed in the appropriate solution for around 16 hours to form required SAM. The samples were then rinsed with water and methanol, followed by drying using a nitrogen gas flow. Three types of samples were prepared: two with different cysteine enantiomers, one with the L enantiomer, and the other with the D enantiomer. The third sample is a racemic mixture consisting of both enantiomers. Gold nanoparticle with optical density 1 with stabilised suspension in citrate buffer solution were purchased from the Merk. If required NPs were deposited on to the SAM functionalised metasurface by immersion into the NP solution, they were rinsed after removal.

***Surface characterisation***: SEM image of the metasurfaces were acquired using SU8240 (Hitachi) at 10 kV.

***Reflection and ORD measurements***: We used a home built polarimeter setup that comprised of tungsten halogen light source (Thor lab) polarisers and 10x objective and a camera. The samples were aligned with camera and reflected light is routed to a spectrometer (Ocean optics USB4000). For reflection spectra plain gold film is taken as background, which are then used for normalisation. ORD spectra were measured using the Stokes method is used. Linearly polarised light is incident at the surface of the sample and reflected light is measured for four different polarisation state (0, ±45, and 90°) with respect to the incident light.

***Raman Measurements***: Raman measurements were performed using NT-MDT NTEGRA Raman microscope. Sample were excited with a 20x Olympus objective lens with a 632.8 nm He-Ne laser with 30 mW power. 5 measurements have been performed on three pairs of enantiomorphic metasurfaces (LH and RH). Each Raman spectrum is acquired for 20 seconds.

***Numerical Simulation:*** Finite element method based numerical simulations were performed using a commercial package (COMSOL V6.0 Mutiphysics Software, Wave Optics module). Single shuriken models with periodic boundary conditions were used to replicate the metasurfaces. A 5 nm chiral dielectric layer is placed over the shuriken to mimic the chiral nature of the adsorbed cysteine. Perfectly matched boundary conditions were used at input and output ports to minimize the internal reflection. The diameter of the shuriken structure is 520 nm with arm width 30 nm, and NPs were just placed above the chiral dielectric layer. To calculate the enhancement and asymmetry average of $|E|^2$ at 632.8 and 703.5 nm in the chiral layer and gap outside the layer is used.

## Acknowledgement

Affar Karimullah would like to acknowledge support by the UKRI, EPSRC (EP/S001514/1 and EP/S029168/1) and the James Watt Nanofabrication Centre. Nikolaj Gadegaard and Malcolm Kadodwala acknowledge support from EPSRC (EP/S012745/1 and EP/S029168/1). Malcolm Kadodwala also acknowledges the Leverhulme Trust (RF-2019-023). Oleksiy Lyutakov would like to acknowledge support by the GACR under the project 20-19353S.

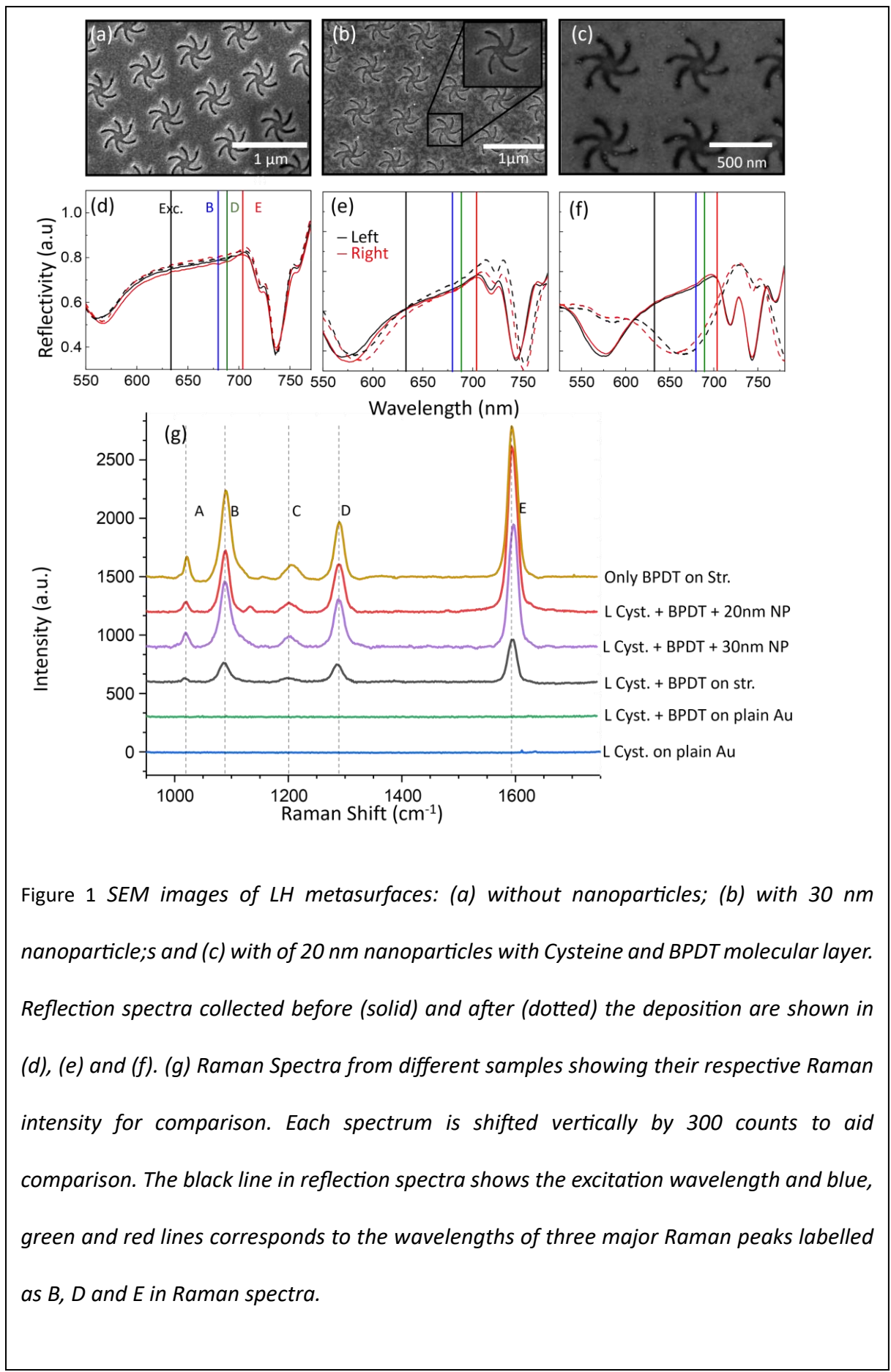

Figure 1 *SEM images of LH metasurfaces: (a) without nanoparticles; (b) with 30 nm nanoparticle;s and (c) with of 20 nm nanoparticles with Cysteine and BPDT molecular layer. Reflection spectra collected before (solid) and after (dotted) the deposition are shown in (d), (e) and (f). (g) Raman Spectra from different samples showing their respective Raman intensity for comparison. Each spectrum is shifted vertically by 300 counts to aid comparison. The black line in reflection spectra shows the excitation wavelength and blue, green and red lines corresponds to the wavelengths of three major Raman peaks labelled as B, D and E in Raman spectra.*

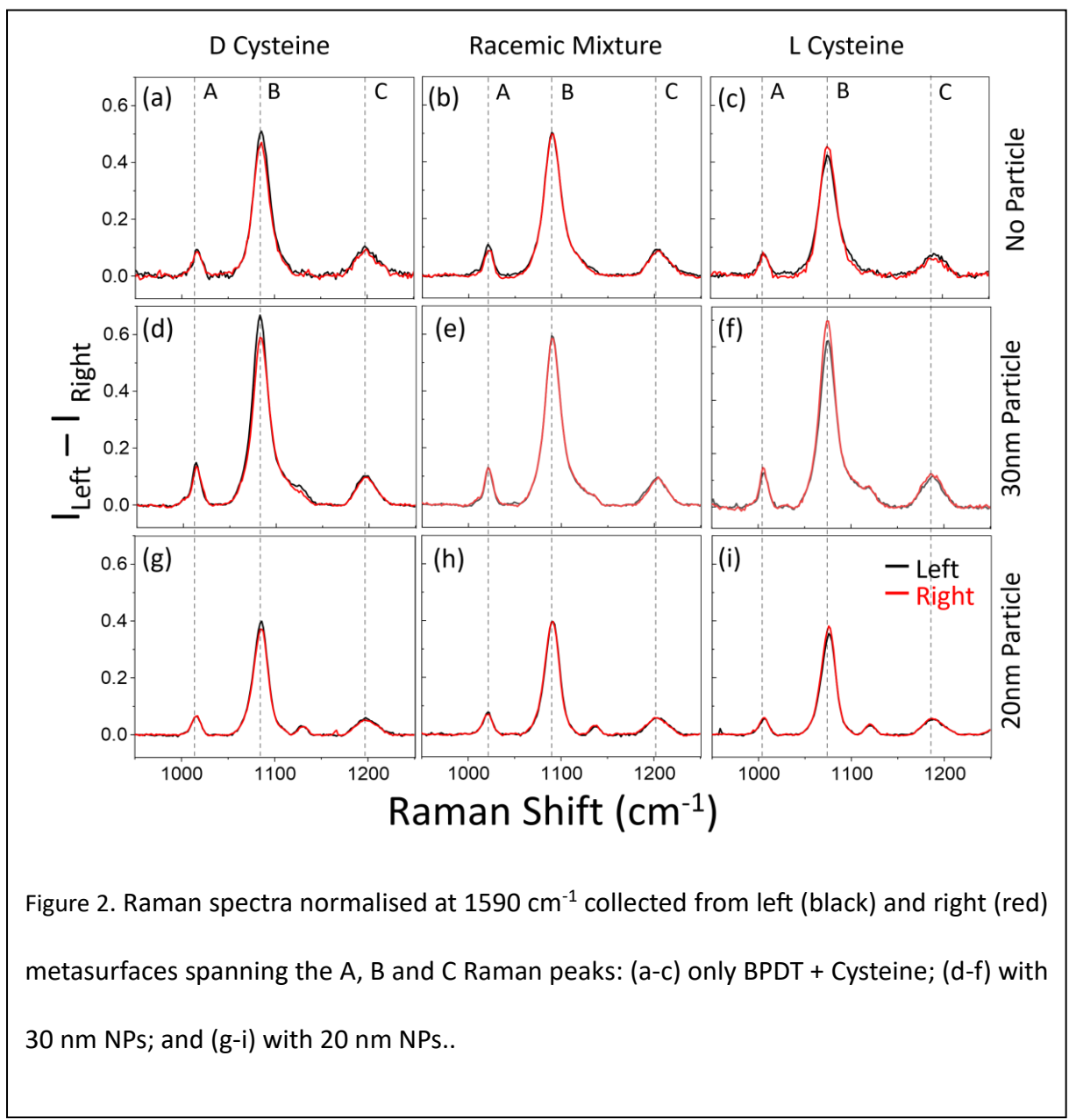

Figure 2. Raman spectra normalised at 1590 cm$^{-1}$ collected from left (black) and right (red) metasurfaces spanning the A, B and C Raman peaks: (a-c) only BPDT + Cysteine; (d-f) with 30 nm NPs; and (g-i) with 20 nm NPs..

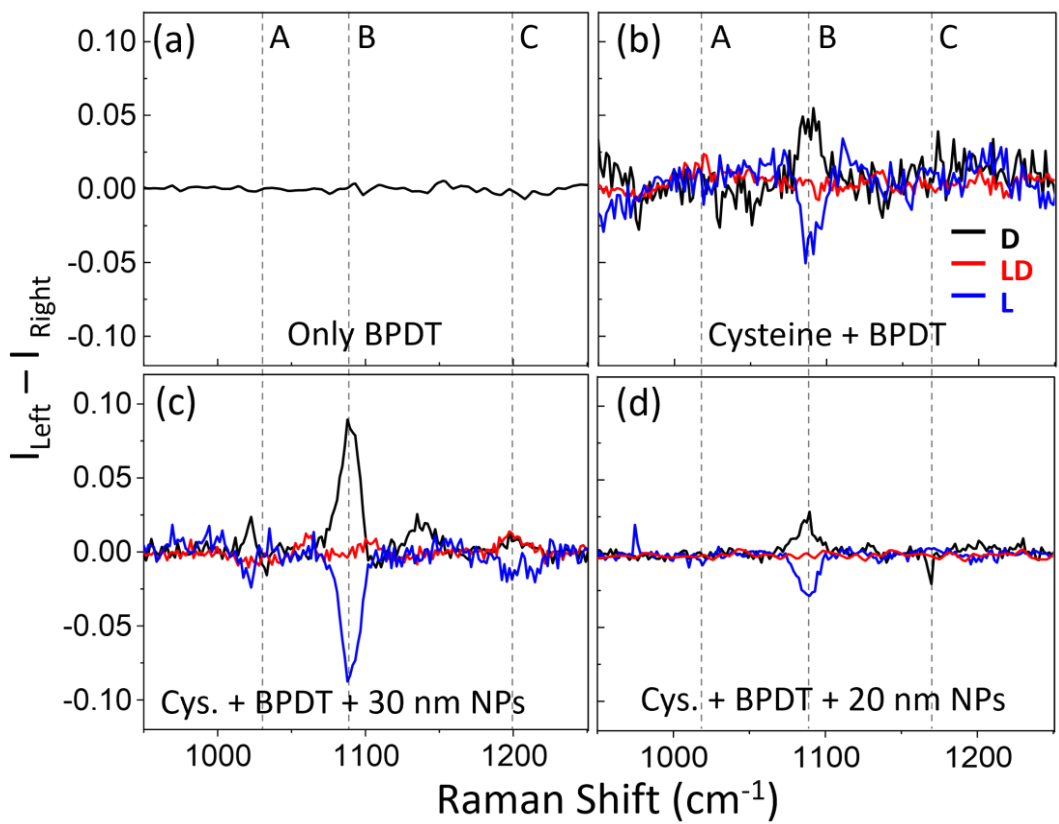

Figure 3. The difference spectra derived from LH minus RH, for: (a) BPDT only. (b) BPDT mixed with cysteine; (c) BPDT with cysteine and 30 nm nanoparticles; (and d) BPDT with cysteine and 20 nm nanoparticles.

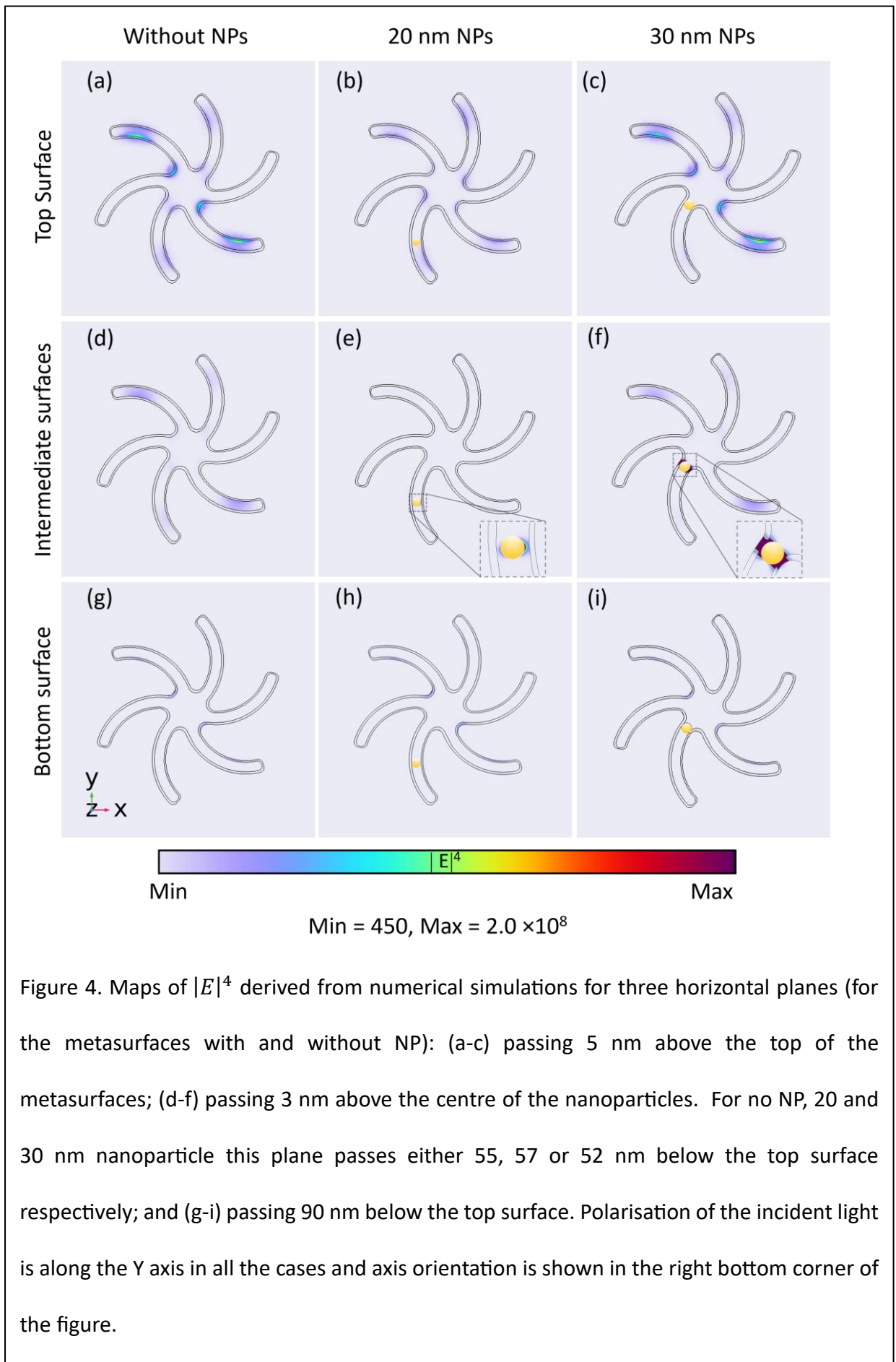

Figure 4. Maps of $|E|^4$ derived from numerical simulations for three horizontal planes (for the metasurfaces with and without NP): (a-c) passing 5 nm above the top of the metasurfaces; (d-f) passing 3 nm above the centre of the nanoparticles. For no NP, 20 and 30 nm nanoparticle this plane passes either 55, 57 or 52 nm below the top surface respectively; and (g-i) passing 90 nm below the top surface. Polarisation of the incident light is along the Y axis in all the cases and axis orientation is shown in the right bottom corner of the figure.

| Raman peak position | Labelled as | Vibrational mode assigned |
|---|---|---|
| 1020 cm$^{-1}$ | A | Ring Deformation |
| 1084 cm$^{-1}$ | B | C$_{ring}$ – S in plane |
| 1200 cm$^{-1}$ | C | 12 C – H in plane |
| 1285 cm$^{-1}$ | D | Inter Ring C – C Streching |
| 1590 cm$^{-1}$ | E | C – C Stretching |

Table 1. Vibrational mode assignments and labelling of BPDT Raman spectra based on reference (18), are tabulated.

| $\lvert I_{Left} - I_{Right} \rvert$ | Without Particle | 30 nm particle | 20 nm particle |
|---|---|---|---|
| Exp. | 0.048 | 0.087 | 0.029 |
| Sim. ($\xi = 1.7 \times 10^{-4}$) | 0.0031 | 0.0063 | 0.0035 |
| Sim. ($\xi = 2.4 \times 10^{-3}$) | 0.066 | 0.117 | 0.082 |

Table 2. Comparison of experimental asymmetries with those derived from numerical simulations, for two $\xi$ values.